\documentclass{article}
\usepackage{mrs2005,epsfig}
\begin{document} 
\title{INTRA-GROUP LIGHT IN HICKSON COMPACT GROUPS - \\ A WAVELET APPROACH}

\author{Cristiano Da Rocha$^{1,2}$, Claudia Mendes de Oliveira$^3$ and Bodo
L. Ziegler$^1$}
\affil{$^1$Institut f\"ur Astrophysik G\"ottingen - Georg August 
Universit\"at G\"ottingen - Germany \\ 
$^2$Divis\~ao de Astrof\'{\i}sica, Instituto Nacional de Pesquisas
Espaciais - DAS/INPE - Brazil \\
$^3$Instituto de Astronomia 
Geof\'{\i}sica e Ci\^encias Atmosf\'ericas - USP - Brazil}

\begin{abstract} 
A diffuse component of intra-group light can be observed in compact groups
of galaxies. This component is presumably due to stellar material tidally
stripped from the member galaxies of the group, which gets trapped in the
group potential. It represents an efficient tool for the determination of
the stage of dynamical evolution of such structures and for mapping the
gravitational potential of the group. Detecting this kind of low surface
brightness structure (about 1\% above the sky level) is a difficult task,
which is subject to many problems like the modeling of stars and galaxy
and sky subtraction. To overcome these problems, we apply a new method,
the wavelet technique OV\_WAV, with which these low surface brightness
structures can be revealed and analyzed by separating different components
according to their spatial characteristic sizes (allowing the study of
the point sources, galaxies and diffuse envelope separately). We have
analyzed 3 of the of the Hickson Compact Groups Catalogue (HCG 79,
HCG 88 and HCG 95) and were able to detect this diffuse component in
two of the studied groups: HCG 79, where the diffuse light corresponds
to $46\pm11\%$ of the total B band luminosity and HCG 95, where the
fraction is $11\pm26\%$. HCG 88 had no component detected, as expected
by its estimated early evolutionary stage.

\end{abstract}

\section{Introduction}
A simple visual inspection of the compact groups of galaxies in Hickson's
catalogue \cite{hic82} reveal that many of them are embedded in a
diffuse intra-group light (IGL) component. This component is a useful
test to measure the intensity of the tidal interactions suffered by the
galaxies and may represent a direct way to map the extension and shape
of the groups gravitational potential and the dark matter halo.

The study of the IGL structures is difficult and complex. With a very low
surface brightness, in general 1 or 2\% above the sky brightness level,
those structures are often dimmed by the group member galaxies. The
analysis process is very sensitive to the sky level estimate and modeling
and subtraction of the galaxies and foreground stars in the field. To
avoid this kind of effect we developed a new method to isolate the IGL
contribution using the ``\`a trous'' wavelet transform, the package
OV\_WAV \cite{epi05}.  The process detects different characteristic
size structures, separating the types of light source in the image and
identifying and measuring the IGL.

\section{Method}
The image analysis was performed with OV\_WAV. The images are decomposed
in wavelet coefficients using the ``\`a trous'' transform which separates
the structures in each scale $n$ with characteristic size, $2^n$
pixels. The detection, identification and reconstruction of objects are
performed using a Multi-Scale Vision Model \cite{bij95}.

With OV\_WAV there is no need of ``a priori'' information about the
sky brightness level and the modeling of bright objects and it is an
optimized method for detection of low surface brightness structures.

\section{Observational Data, Analysis and Results} 
We have studied a sample of 3 Hickson compact groups \cite{hic82} (HCG
79, HCG 88 and HCG 95) as a pilot study for a larger survey to search
for IGL. The groups were chosen for being in different evolutionary
stages. Deep B and R images, obtained at the CFHT, were used in this
analysis.

The detection and analysis of the IGL component of each group was
performed with OV\_WAV and simulated images were also analyzed in order to
determine OV\_WAV configuration parameters and estimate detection limits
and systematic errors. The simulations showed that we are able to detect
low-surface brightness extended structures, down to a $S/N = 0.1$ per
pixel, which corresponds to a 5-$\sigma$-detection level in wavelet space.

\begin{figure}  
\begin{center}
\epsfig{figure=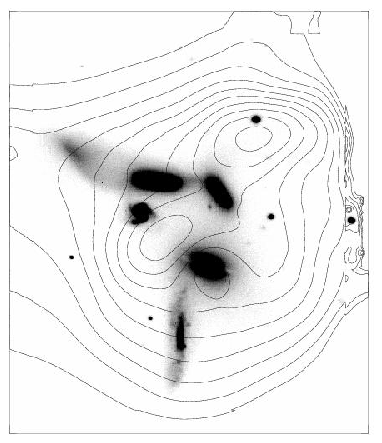,width=5.5cm}  
\epsfig{figure=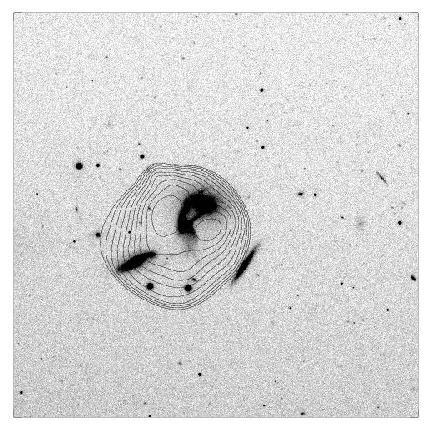,width=5.7cm}  
\end{center}
\caption{The left panel shows the B band image of HCG 79 with IGL in
contour curves superposed, ranging from $24.2$ to $25.1$ magnitudes in
steps of $0.1~B~mag~arcsec^{-2}$. The right panel shows the B band image
of HCG 95 with IGL in contour curves superposed, ranging from $26.9$
to $27.8$ magnitudes in steps of $0.1~B~mag~arcsec^{-2}$. 
\label{figdif}} 
\end{figure}

Irregular IGL components were detected in 2 groups (HCG 79 and HCG 95)
and no component was detected in HCG 88 up to the detection limit
($29.1~B~mag~arcsec^{-2}$). The detected IGL components are shown
in figure~\ref{figdif} and the main properties are summarized in
table~\ref{tabres}.

\begin{table}

\centering
\caption{Properties of the IGL component detected in our sample.
\label{tabres}}
\begin{tabular}{llllllll}

\hline
Group & \multicolumn{2}{c}{\% ($B$ and $R$)} & \multicolumn{2}{c}{$<\mu>$ ($B$ and $R$)} & \multicolumn{2}{c}{Mag. ($B$ and $R$)} & $(B-R)_0$ \\
\hline

HCG 79  & $46\pm11$\% & $33\pm11$\% & $24.8\pm0.16$ & $23.9\pm0.16$ & $14.0\pm0.16$ & $13.1\pm0.16$ & $0.86\pm0.22$ \\
HCG 95  & $11\pm26$\% & $12\pm10$\% & $27.3\pm0.30$ & $25.5\pm0.15$ & $16.9\pm0.30$ & $15.1\pm0.15$ & $1.75\pm0.34$ \\
\hline
\end{tabular}

\end{table}

%
%

\section{Conclusions and Perspectives} 
We have detected an IGL component in HCG 79 and HCG 95, groups that
present clear signs of galaxy interaction, indicating a compact and bound
configuration for a few crossing times. HCG 79 presents an irregular IGL
distribution, coherent with the X-Ray distribution, following the group
potential, as well as the hot gas \cite{pil95} and this component is bluer
than the galaxies ((B-R) = 1.5), because it may be a mix of two different
sources: stripped material from only the outer parts of the galaxies and
the destruction of relatively blue dwarf galaxies. HCG 95 has an almost
spherical IGL distribution around the interacting system HCG 95A/C,
with colors that are coherent with the galaxies' red colors indicating
an old stellar population.  The absence of an IGL component in HCG 88
indicates an early stage of dynamical evolution, as expected by other
group properties like the absence of strong signs of galaxy interaction.

Therefore, we can envision an evolutionary sequence for our sample,
with HCG 79 being the most evolved group and in an advanced stage of
dynamical evolution, with HCG 95 being in an intermediate stage and HCG
88 in an initial epoch still without an IGL components.

The presence of an IGL component indicates gravitationally bound
configurations in which tidal encounters already stripped a considerable
fraction of mass from the member galaxies and an advanced stage of
dynamical evolution, providing tests for formation and evolution models
of groups.

We are now conducting a survey to search for IGL in more HCG. We then
will compare the measured light fractions and shapes to predictions
from N-Body simulations in order to assess the dynamical age of those
dense structures. We have already obtained data with the Laica Wide
Field Camera at Calar Alto (Spain), currently being processed, and were
awarded time at the Wide Field Camera of INT (La Palma) in November 2005.

\acknowledgements{
This project is supported by FAPESP grants No. 96/08986-5 and 02/06881-4,
CAPES/DAAD grant No. BEX: 1380/04-4 and VW Junior Research Group -
Kinematic Evolution of Galaxies - Volkswagen Foundation (I/76 520)}

\vfill 

\begin{thebibliography}{}{ 
\bibitem{bij95} Bijaoui, A. \& Ru\'e, F. 1995, Signal Processing, 46, 345
\bibitem{epi05} Epit\'acio Pereira, D. N., Raba\c ca, C. R. \& Da Rocha, C. 
2005, in preparation
\bibitem{hic82} Hickson, P. 1982, ApJ, 255, 382
\bibitem{pil95} Pildis, R. A., Bregman, J. N. \& Evrard, A. E. 1995, ApJ, 443, 
514
 
} 
\end{thebibliography}
\end{document}